\documentclass[twocolumn,prl,showpacs,floatfix]{revtex4}
\usepackage{graphicx}
\usepackage{epsfig}
\usepackage{rotating}
\usepackage{amsmath}
\usepackage{amsfonts}
\usepackage{amssymb}
\usepackage{enumerate}
\usepackage{longtable}
\usepackage{dcolumn}
\usepackage{bm}
\newcommand{\chiSG}{\chi_{\raisebox{-0.6ex}{$\scriptstyle SG$}}}
\begin{document}

\title{de Almeida-Thouless instability in short-range Ising spin-glasses}

\author{R. R. P. Singh}
\affiliation{University of California Davis, CA 95616, USA}

\author{A. P. Young}
\affiliation{University of California Santa Cruz, CA 95064, USA}

\date{\rm\today}

\begin{abstract}
We use high temperature series expansions to study the $\pm J$ Ising
spin-glass in a magnetic field in $d$-dimensional hypercubic lattices
for $d=5, 6, 7$ and $8$, and in
the infinite-range Sherrington-Kirkpatrick (SK) model. The expansions are obtained in the
variable $w=\tanh^2{J/T}$ for arbitrary values of $u=\tanh^2{h/T}$ complete to
order $w^{10}$.  We find that the scaling dimension $\Delta$ associated with
the ordering-field $h^2$ equals $2$ in the SK model and for $d\ge 6$. However,
in agreement with the work of Fisher and Sompolinsky, there is a violation of
scaling in a finite field, leading to an anomalous $h$-$T$ dependence of the
Almeida-Thouless (AT) line in high dimensions, while scaling is restored as $d
\to 6$. Within the convergence of our series analysis, we present evidence
supporting an AT line in $d\ge 6$. In $d=5$, the exponents $\gamma$ and
$\Delta$ are substantially larger than mean-field values, but we do not see
clear evidence for the AT line in $d=5$.
\end{abstract}


\maketitle


One of the most striking predictions of the mean field theory of spin
glasses~\cite{edwards:75,sherrington:75,binder:86} is the existence of a line
of transitions in the magnetic-field temperature plane first found by de
Almeida and Thouless (AT)~\cite{almeida:78}. This transition is
surprising since it occurs without
the breaking of any ``obvious" symmetry, and instead marks the onset of
``replica symmetry breaking'' (RSB). The solution of the mean-field,
infinite-range,
Sherrington-Kirkpatrick (SK)~\cite{sherrington:75} model in the RSB phase
below the AT line is complicated and 
was obtained, in a tour-de-force,
by Parisi~\cite{parisi:80,parisi:83}. Whether or
not an (AT) line of instabilities occurs in the magnetic-field temperature
plane for \textit{short-range}
Ising spin-glasses has been an outstanding problem that
has remained unresolved over the last few decades. According to the ``droplet
theory''~\cite{fisher:87,fisher:88} of spin glasses, the AT line is an
artefact of the infinite-range nature of the interactions of the SK model and
would not occur in \textit{any} finite-dimensional, short-range model. In the
``RSB scenario", see e.g.~\cite{baity-jesietal:14,banosetal:12} and references
therein, the behavior of real spin glasses is similar to that of the SK model
and so there would be an AT line down to $d=3$.

In recent years, numerical simulations have investigated whether or not there
is an AT line in short-range spin glasses. While some work finds evidence for
an AT line only above dimension $d=6$~\cite{katzgraber:05,larson:13}, other
work~\cite{baity-jesietal:14,banosetal:12}, which used a different method
doing finite-size scaling, does find an AT line at least down to $d=4$ and
possibly $d=3$.

In view of the importance of whether or not an AT line exists in short-range
spin glasses,
and the fact that Monte Carlo
simulations don't give an unambiguous answer, it is useful to attack the
problem by
other possible means, and here we use high-temperature series expansions.
Some benefits
of the series methods are that averaging over disorder is done exactly, it can
be done in arbitrary dimension, is particularly accurate in high dimensions,
and that the series is an equilibrium property of the \textit{infinite}
system. In fact, Monte Carlo studies of the AT line in the range of dimension
that we consider here, $d \ge 5$, have not been performed directly, but only
\textit{indirectly} using a one-dimensional model with long-range
interactions~\cite{katzgraber:05,leuzzi:08}.

Obtaining the series in a field is more complicated than in zero field, so we
are only able to obtain series of moderate length. Nonetheless, the series
does provide evidence for an AT line above $d=6$. Below $d=6$ the series does
not find good evidence for an AT line, but whether this is because the line
does not exist or the series is not long enough to see it, is unclear.
This situation is reminiscent of an early perturbative renormalization group
calculation~\cite{bray:80b} which did not find a stable fixed point in a field
below $d=6$. In that case, the issue, as yet unresolved, is whether the AT
line does not exist below $d=6$ or whether there is one which is just not
accessible by perturbation theory.



We consider the Hamiltonian
\begin{equation}
\mathcal{H} = -\sum_{\langle i, j\rangle} J_{ij} S_i S_j - h\sum_{i=1}^N S_i
\, ,
\end{equation}
where the $S_i$ are Ising spins which take values $\pm 1$, and the
interactions $J_{ij}$ are \textit{quenched} random variables with a bimodal
distribution, i.e. $J_{ij} = \pm J$ with equal probability.
The $N$ spins either lie on a hypercubic lattice, in which
case the
interactions are between nearest-neighbors and have $J=1$,
or correspond to the
Sherrington-Kirkpatrick (SK)~\cite{sherrington:75}
model in which case there is no lattice
structure, every spin interacts with every other spin, and 
$J = 1/\sqrt{N}$. We choose a bimodal distribution because the series can be
worked out much more efficiently for this case than for a general
distribution~\cite{SY:17-TBP}.

The AT line is characterized by the divergence of the spin glass
susceptibility $\chiSG$ where
\begin{equation}
\chiSG = {1\over N} \sum_{i, j= 1}^N \left[ \bigl(\,\langle S_i S_j \rangle -
\langle S_i \rangle\, \langle S_j \rangle\,\bigr)^2 \right]_\mathrm{av} \, ,
\end{equation}
where $[\cdots]_\mathrm{av}$ denotes an average over the disorder. 
For a fixed value of $h$ we expand the susceptibility for the hypercubic
lattice in powers of
\begin{equation}
w = \tanh^2(J/T) \,.
\label{w_def}
\end{equation}
The coefficient of $w^n$ turns out to be a polynomial of order $2n +2$ in
\begin{equation}
u = \tanh^2(h/T),
\end{equation}
so
\begin{equation}
\chiSG(w, u) =\sum_{n=0}^\infty \,\left(\,  \sum_{m=0}^{2n+2} a_{n,m} u^m
\,\right)\, w^n \, .
\label{series}
\end{equation}
We evaluate all the coefficients $a_{n, m}$ up to  order $n=10$ for a
hypercubic lattice in arbitrary dimension $d$~\cite{SY:17supp_mat}. 
The series for the SK model is
obtained by setting $J=1/{2d}$ 
and taking $d \to \infty$ limit, in which case
the high temperature expansion variable becomes $x=(1/T)^2$ rather than
$w$. 
A ten term
series is only of moderate length, but, compared with zero field, determining
the series in a field requires much more computer time and memory, so it would
need a very large numerical effort to \textit{substantially} increase the
number of terms beyond 10. 
Part of the reason for the extra complexity
is that the coefficients rapidly become large,
and, for a given order $n$, considerable cancellations occur between the
coefficients with different values of $m$. For example, for the SK model with
$u=0.2$ the largest individual
contribution to the coefficient of $w^{10}$ is about $10^7$
times greater than the final answer~\cite{SY:17supp_mat}.

Equation~\eqref{series} gives the high temperature series expansion for
\textit{arbitrary} values of the ratio $h/T$. Fixing this ratio corresponds to
expanding $\chiSG$ along a diagonal line in the $h$-$T$ plane ending at
$h=T= 0$, which \textit{must} intersect an AT line if one exists.



{\it Zero-field Spin-Glass Transition}:
It is known from earlier
studies~\cite{fisch:77,singh:86,klein:91,daboul:04},
that a $10$-term series in $w$ does not give
consistent indication of a critical temperature in $d=3$ and is also
problematic in $d=4$ giving rather large and inconsistent values of critical
exponents. Hence, we confine our
analysis to $d=5$ and higher. 

For the SK model,
the zero-field spin-glass series is a simple geometrical series, 
in $x = 1/T^2$,
which sums to ${1/(1-x)}$ showing that the exponent $\gamma$ equals unity.
For $d\ge 6$, the model's critical behavior is governed by a Gaussian fixed
point with $\gamma=1$~\cite{harris:76,chen:77}.
We use 
Pad\'e approximants directly on the series
to estimate $w_c$, the critical value of $w$. This fixes 
$\gamma$ to unity, and produces estimates for $w_c$ shown
in Table~\ref{table}. Note that uncertainties in series analysis are just confidence limits~\cite{oitmaa:06}.
These $w_c$-values are consistent with those from large dimensionality
expansions~\cite{singh:88,fisher:90}, and agree to within around a percent with those from
a more sophisticated analysis taking
corrections to scaling into account~\cite{klein:91,daboul:04}.  
It is well known that estimates of
critical points in series-analysis are correlated with estimates of
critical exponents. Hence, by fixing the critical exponent to unity, we avoid
some of the subtleties and get a fairly reasonable estimate of the critical
point with a moderate length series.

\begin{table}[h]
\caption{\label{table}Estimates of critical points and exponents in various
dimensions for zero field. 
}
\begin{tabular}{|c|c|r|r|}
\hline
\ d\   & $w_c \ \ \ \ $ & $\gamma$ & $\Delta$ \\
\hline\hline
8  & $\ 0.0695\pm 0.0002$ \ & 1 & $2.0\pm 0.1$ \\
7  & $0.0816\pm 0.0004$ & 1 & $2.0\pm 0.2$ \\
6  & $0.0996\pm 0.0008$ & 1 & $2.0\pm 0.5$ \\
5  & $0.1388\pm 0.0009$ & $1.9\pm0.1$ & $3.1\pm 0.4$\\
\hline
\end{tabular}
\end{table}


In $d=5$, we use standard d-log Pad\'e approximants and differential
approximants~\cite{hunter:79,fisher:79} to analyze the series. The critical
point estimate $w_c=0.1388\pm 0.0009$
is consistent with previous studies~\cite{fisch:77,klein:91}. 
Using biased approximants with critical point fixed at the central estimate $w_c=0.1388$,
we obtain $\gamma\ =\  1.9\pm 0.1$, again in agreement with previous studies.

{\it Scaling dimension of the ordering field:}
Field theory predicts~\cite{harris:76,chen:77,fisher:85} that the scaling dimension of the
ordering field $h^2$, or equivalently $u$, should be $\Delta=2$ at the Gaussian
fixed point. In other words for $d\ge 6$, the variable $u$ should scale, near
$T_c$, with the reduced temperature
$t\equiv(T-T_c)/ T_c$, in the combination $u/t^2$.

To study this through series expansions we consider two single variable series
in $w$ defined as
\begin{equation}
K_1(w) = {\displaystyle 
\partial_u \chiSG(w, u) |_{u=0}
\over \chiSG(w, 0)}\, ,
\label{K_1}
\end{equation}
and
\begin{equation}
K_2(w) = {
\partial^2_u \chiSG(w, u) |_{u=0}
\over
\partial_u \chiSG(w, u) |_{u=0} } \, .
\label{K_2}
\end{equation}
Both quantities $K_1(w)$ and $K_2(w)$
should diverge at the critical temperature as
$1/t^\Delta$ with $\Delta = 2$ for $d \ge 6$. Note that we consider the limit
$t\to 0$ for which $t \equiv (T-T_c)/T_c \propto (w_c - w)/w_c$.

For the SK model, 
these quantities sum up to 
\begin{equation}
K_1^{SK}={-2/ (1-x)^2}
\end{equation}
and
\begin{equation}
K_2^{SK}=6-2x-{(x+7)/ (1-x)^2}
\end{equation}
respectively, with $x=1/T^2$,
clearly showing that $\Delta=2$.  In fact,
from an asymptotic analysis of our graphical method, one can show that
the $m$-th derivative of $\chiSG^{SK}$ with respect to $u$, evaluated at
$u=0$ diverges as $1/t^{1+2m}$, confirming that $\Delta=2$ is true to all
perturbative orders in $u$.

\begin{figure}
\begin{center}
\includegraphics[angle=270,width=0.47\columnwidth]{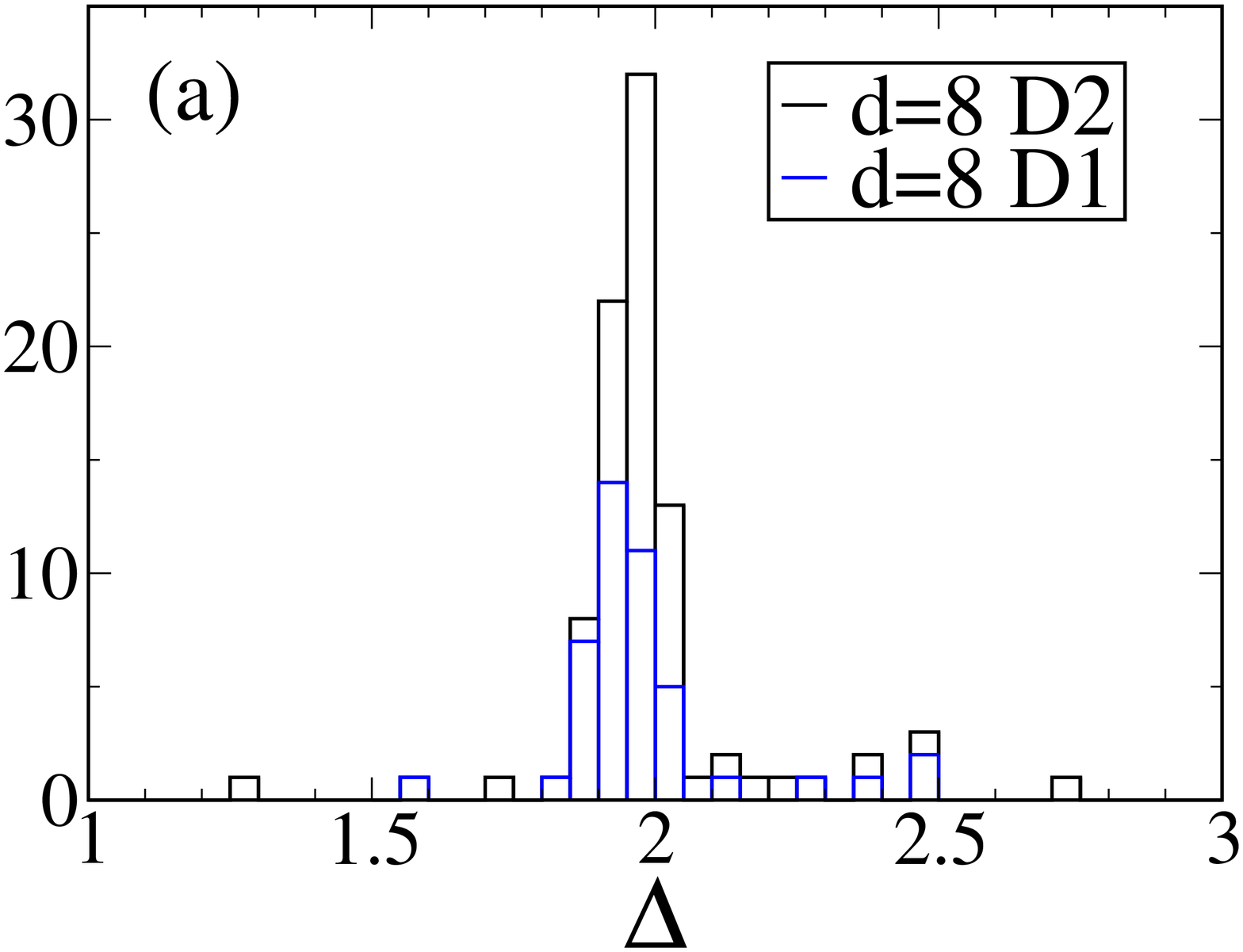}
\includegraphics[angle=270,width=0.47\columnwidth]{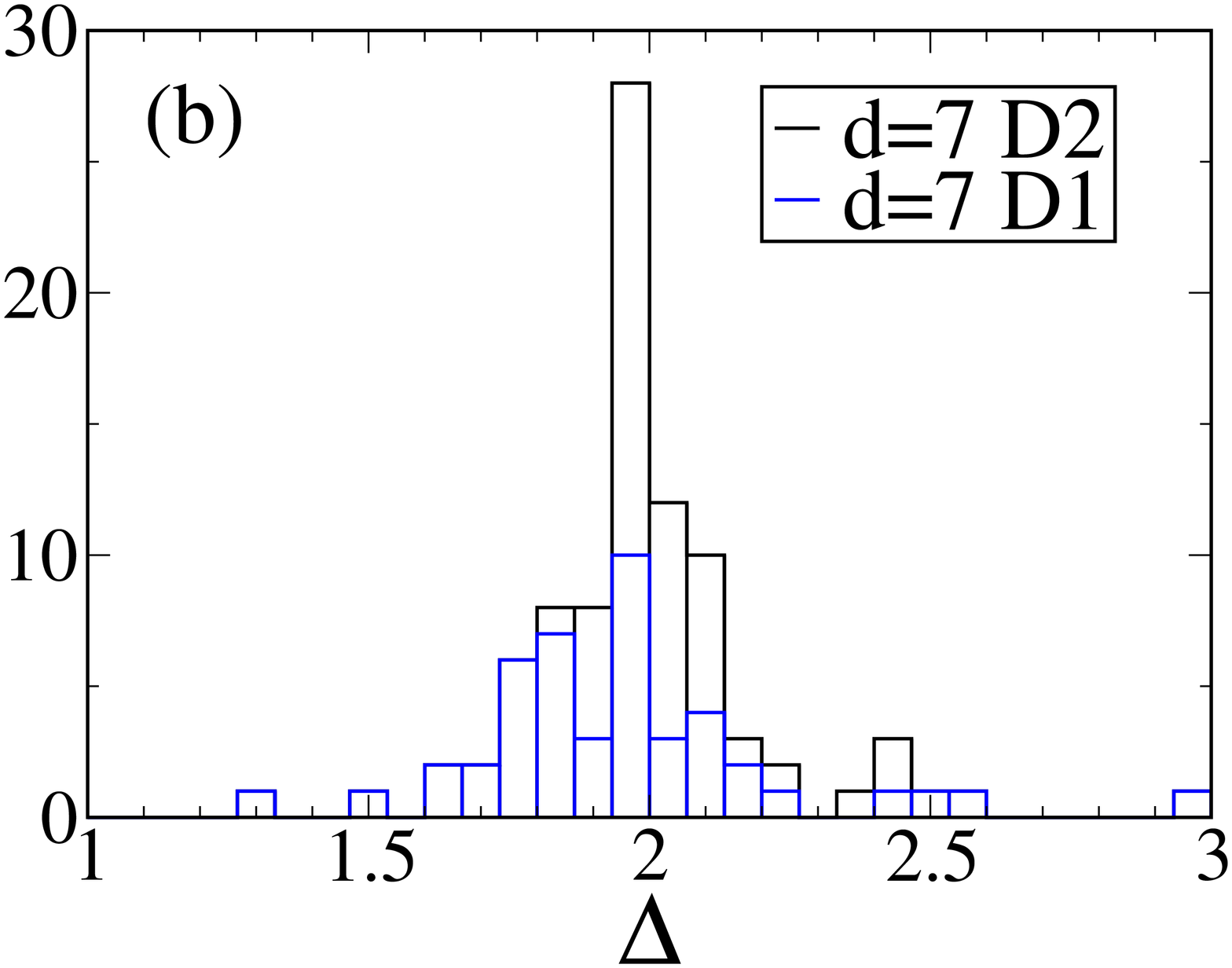}

\includegraphics[angle=270,width=0.47\columnwidth]{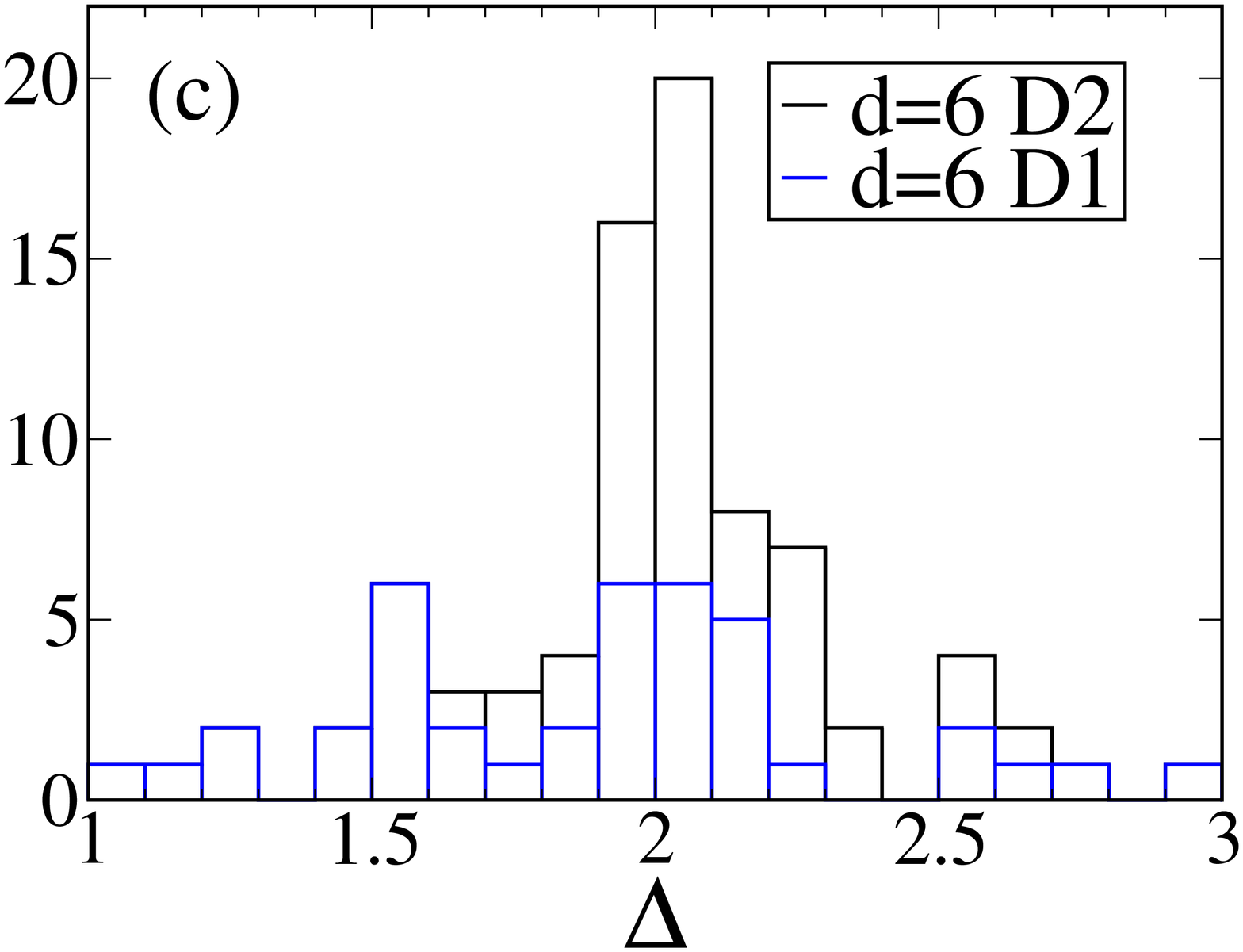}
\includegraphics[angle=270,width =0.47\columnwidth]{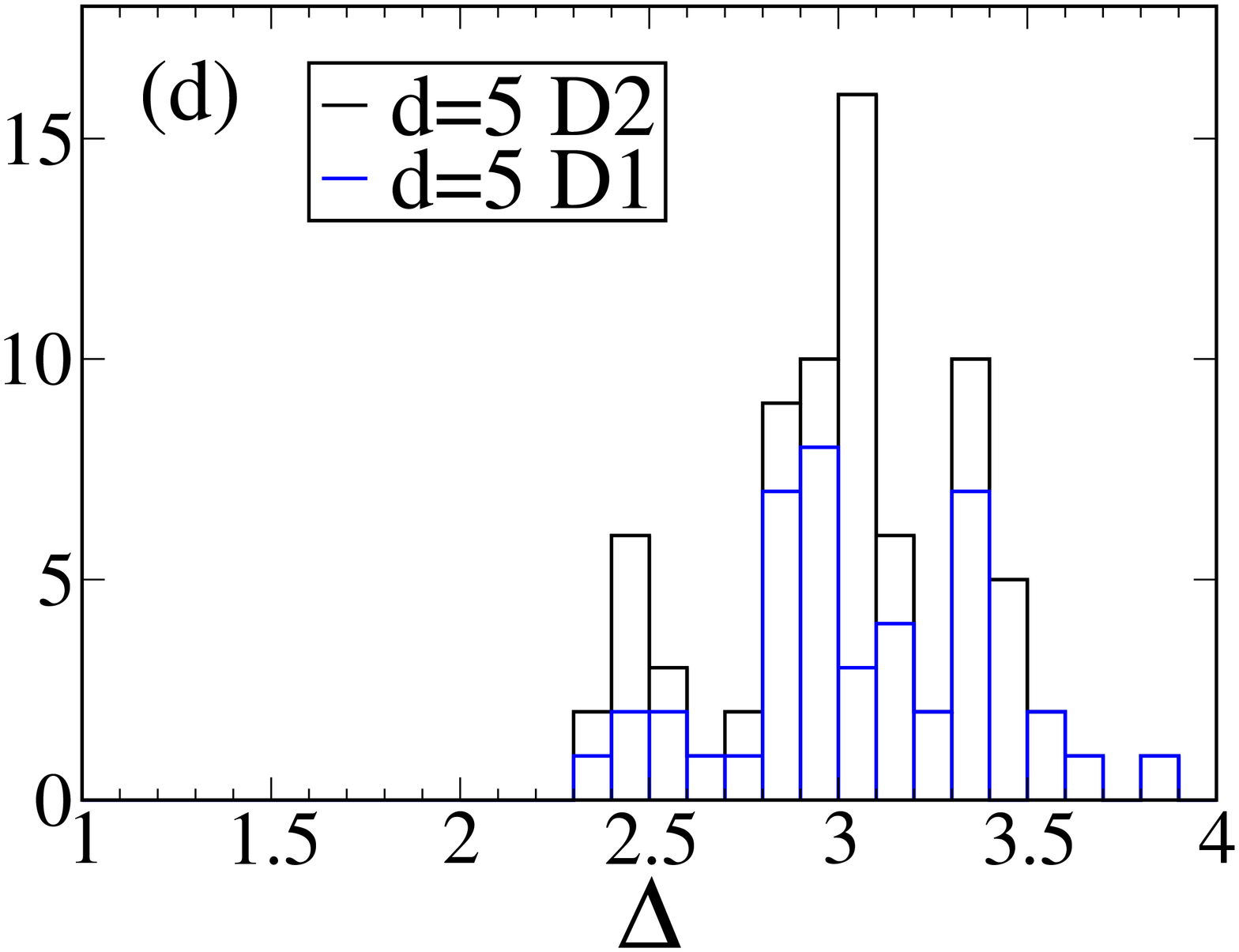}

\caption{Histogram of the exponent $\Delta$ in zero field obtained from various approximants
in (a) $d=8$, (b) $d=7$, (c) $d=6$ and (d) $d=5$ to the series for
$K_1 (D_1)$ and $K_2 (D_2)$ defined in Eqs.~\eqref{K_1} and \eqref{K_2}.
\label{hist_Delta}}
\end{center}
\end{figure}

To analyze $K_1(w)$ and $K_2(w)$ series in finite dimensions, we use d-log
Pad\'es and differential approximants. We fix the critical point at those
values estimated from the zero-field susceptibility series, see
Table~\ref{table}.  A histogram of $\Delta$ values estimated from the analysis
are shown in Fig.~\ref{hist_Delta}. It is clear that in $d\ge6$ the exponent
$\Delta$ remains equal to $2$. However, in $d=5$ it is closer
to $3$.

{\it Analysis of series in a finite field:}
We fix a value of $u$ and study the series in $w$.
In the SK model
the spin-glass susceptibility diverges as a simple pole at the AT line. Unlike
the case of zero field 
the series for SK model are no longer simple in a field, and indeed no
truncated series can reproduce exactly the violation of scaling encapsulated
in the fact that, along the AT line, $T-T_c$ scales as $h^\theta$ with
$\theta=2/3$ rather than as $h^{2/\Delta}$ as expected from scaling.  In fact,
for any finite length series, at sufficiently small $h$, such a non-linear
relation can not follow. Hence, our
focus will be on fields which are not too small to be dominated by just
the leading order field terms.

\begin{figure}
\begin{center}
 \includegraphics[width=0.47\columnwidth]{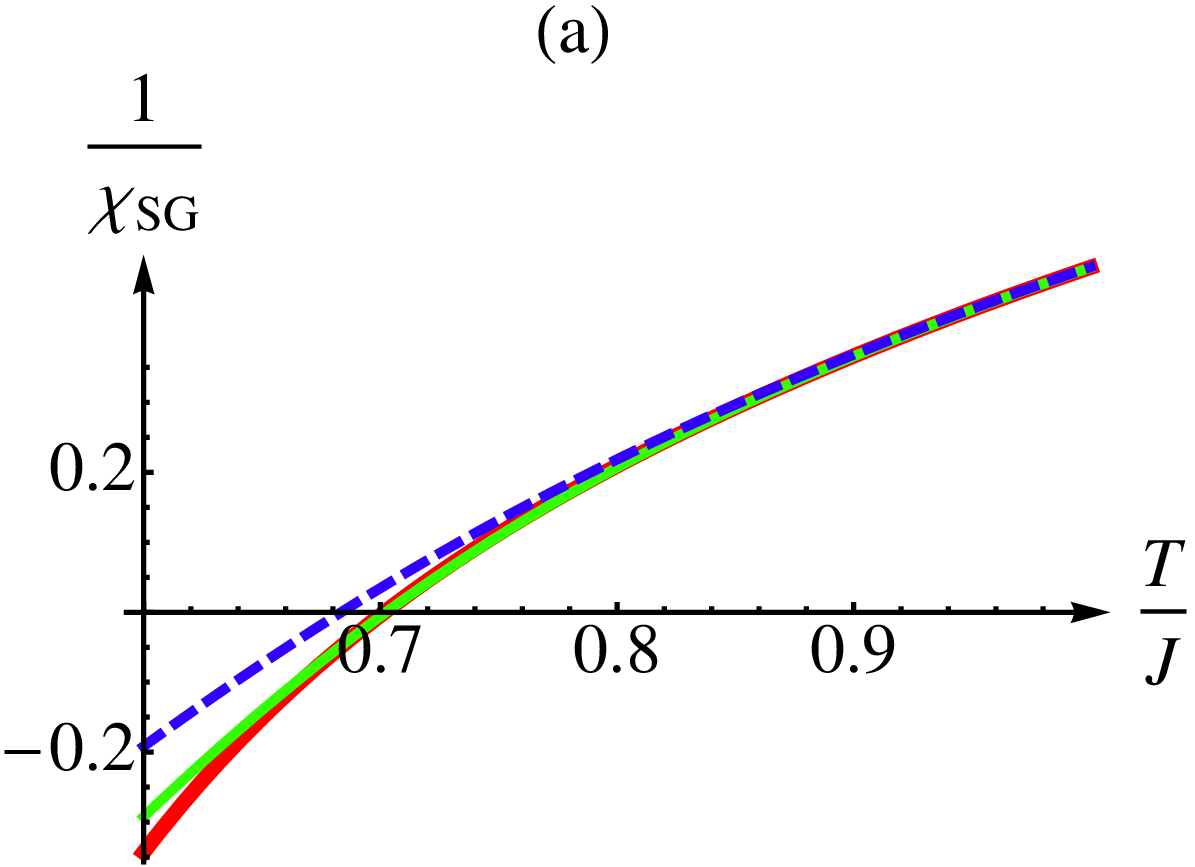}
 \includegraphics[width=0.47\columnwidth]{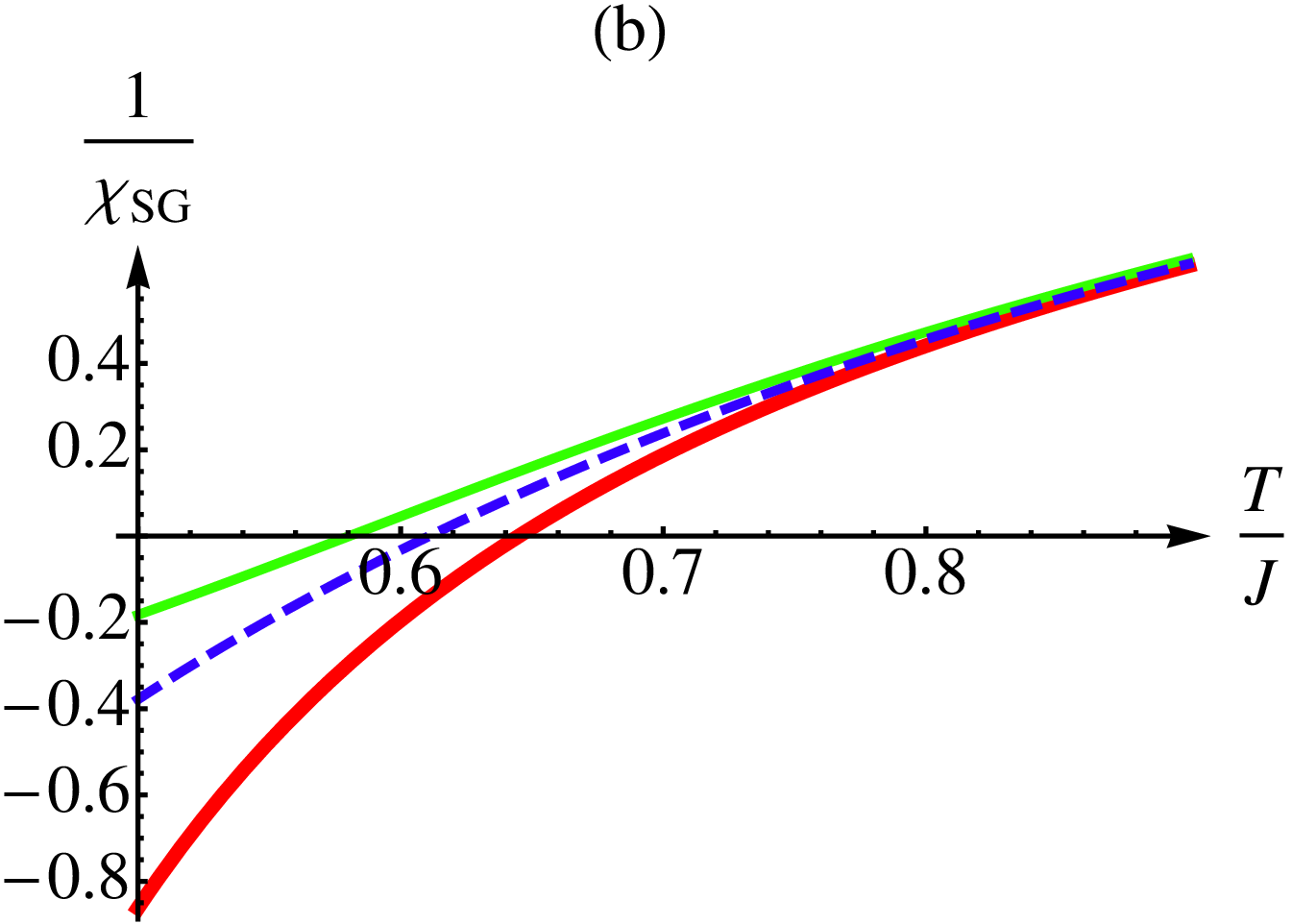}
\caption{The inverse of the spin glass
susceptibility in the SK model for (a) $u \equiv \tanh^2(h/T) = 0.1$ and
(b) $u = 0.2$.
The AT line
is where $\chi_{SG}^{-1} =0$.
The exact values obtained numerically from
the equations of AT are shown by the solid red (thick) 
line. Also shown are the (5, 5) (green, solid) and (4, 4) (blue,
dashed)
Pad\'e approximants of the
high temperature series using the transformed variable $w = \tanh^2(1/T)$.
It is seen that the divergence is located reasonably well at $u=0.1$
but less so at $u=0.2$.
\label{chisg_SK}}
\end{center}
\end{figure}

We have found that the finite-field series for the SK model do not converge well
close to the AT line. 
The series analysis works better in the variable
$w=\tanh^2{1/T}$.  Two diagonal Pad\'e approximants for 
$\chi_{SG}^{-1}$
using the variable $w$ for $u \equiv \tanh^2(h/T) =0.1$ and $u=0.2$
are shown in
Fig.~\ref{chisg_SK} along with the exact value computed numerically.  
The critical point is located reasonably well at $u=0.1$
but not at $u=0.2$. This is found to be true for a majority of
Pad\'es, including off-diagonal ones.

\begin{figure}
\begin{center}
 \includegraphics[angle=270,width=\columnwidth]{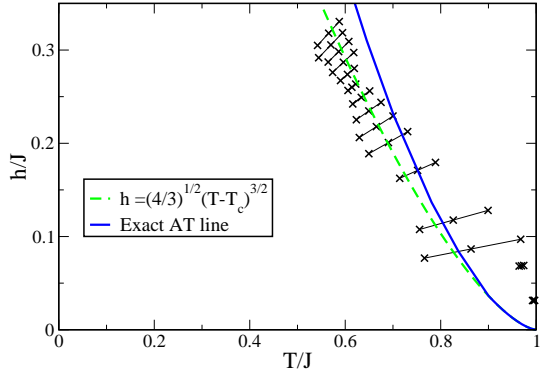}
\caption{\label{ATline_SK} 
The solid (blue) line is the expression for the AT line in the SK model 
while the dashed (green) line is its asymptotic small $h$ limit. The points
joined by lines are the results of the Pad\'e analysis of the 10-term series,
including just the Pad\'e approximants that are bunched together.
}
\end{center}
\end{figure}

For different values of $u$, we carry out a large number of Pad\'e
approximants and determine the critical point from the set of approximants
which are bunched closest to each other. The estimated phase boundary is shown
in Fig.~\ref{ATline_SK}.  The exact value of AT line for the SK model,
determined numerically, and its asymptotic small $h$-$t$ limit are also shown in the figure.  
The series analysis is consistent with the correct $\theta=2/3$ value for the AT line but 
overestimates the extent of the paramagnetic phase for larger $u$. Any significant improvement will
need substantially longer series.
Note that as discussed in the previous paragraph, for very small $h$, the analysis is dominated
by the leading $h$-terms and shows only a small shift in the critical point.
While the convergence is not excellent, it is clear that high temperature expansion with a moderate
number of terms can capture the highly non-trivial Almeida-Thouless
instability in the SK model. 

\begin{figure}
\begin{center}
 \includegraphics[angle=270,width=\columnwidth]{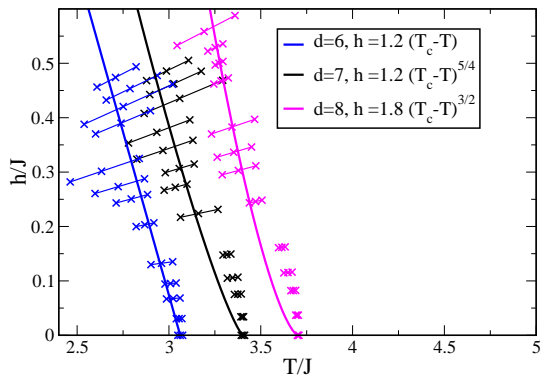}
\caption{\label{ATline_sr} 
Estimates of the AT line in $d=6, d=7$ and $d=8$ obtained from Pad\'e analyses
of the series. The formulae for the lines are given
in the legend and are discussed in the text. 
}
\end{center}
\end{figure}

\begin{figure}
\begin{center}
 \includegraphics[angle=270,width=\columnwidth]{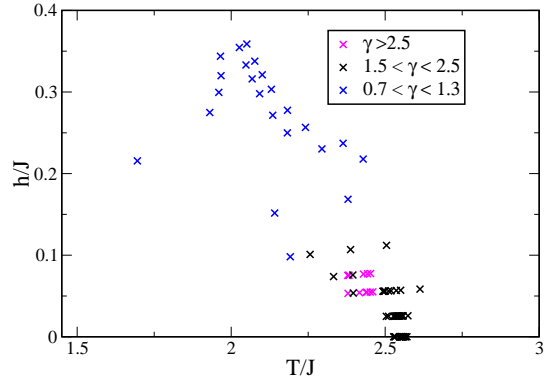}
\caption{
Approximants showing divergence of the spin-glass susceptibility in $d=5$,
and the range of their exponent $\gamma$.
\label{div_5d}}
\end{center}
\end{figure}

Fisher and Sompolinsky~\cite{fisher:85}
have shown that between $d=8$ and $d=6$ the AT-line
exponent becomes $\theta={4 / (d-2)}$, which goes from the SK value of
$\theta=2/3$ in $d=8$ to $\theta=1$ in $d=6$. Thus usual scaling relation
$\theta=2/\Delta$ is restored in $d=6$.  For $d=6,\ 7,\ 8$, we repeat the same
analysis as for the SK model. The results for the estimated AT lines are shown
in Fig.~\ref{ATline_sr}. The uncertainties in locating the AT line are too large to
allow an unbiased fit to a power-law. However, a few points can clearly be
noted: (i) In both $d=7$ and $d=8$, the small-field behavior differs
qualitatively from that at larger fields and is very similar to the behavior
seen in the SK model. (ii) For $d=7$ and $8$ it
is only after $u$ exceeds a certain value that a
more consistent behavior with $\theta<1$ emerges. As a guide, we have drawn
curves with $\theta=2/3$ and $\theta=4/5$ in $d=8$ and $d=7$ respectively, as
expected from the analysis of Fisher and Sompolinsky. (iii) In $d=6$, we do
not see the clear discrepancy at low-fields and the behavior is more
consistent with $\theta=1$ as expected from scaling, which is predicted to be
restored~\cite{fisher:85} in $d=6$. This suggests that the series analysis is
capturing some of the key features of the finite-field behavior in
short-range spin-glasses and its changes with dimensionality, and that the
Almeida-Thouless instability does exist for $d\ge 6$.

In $d=5$, the system is no longer governed by a Gaussian fixed point. However,
the increased $\Delta$ value shown in Table~\ref{table} suggests that if there
is an AT line it should have
a power $\theta = 2 / \Delta$ which is again close to $2/3$. For this case, we
analyze our series by d-log Pad\'e approximants and differential approximants.
At very small $u$ values, we see a singularity that is very similar to what is
seen in the SK model and in higher $d$. There is only a small shift in the
critical point.  But, once $u$ exceeds a certain value most
approximants do not show a consistent divergence. As seen in Fig.~\ref{div_5d},
only a handful of approximants show any divergence at all. These predict a
reduced exponent $0.7<\gamma<1.3$.
This could imply that the series are too short to
see the non-trivial critical behavior in $d<6$ or it could mean that there is
no Almeida-Thouless instability below $d=6$. 
We especially note that one difference
in our analysis of the susceptibility in a field in $d=5$ versus higher-$d$ is that in higher dimensions we biased
the critical exponents to have mean-field values. The absence of such a bias contributes to the
uncertainty in the $d=5$ analysis and may be partly responsible for the lack of a more definitive
answer in $d=5$.


{\it Conclusions:} In conclusion,
we have studied the problem of short-range Ising spin-glasses in a field by
high temperature series expansion methods. We have presented evidence for
violation of scaling along the AT line in high dimension and its restoration
as $d\to 6$, as first shown by Fisher and Sompolinsky~\cite{fisher:85}.
Within the convergence of our analysis, we have presented
evidence for the existence of the AT line of instabilities for $d\ge 6$. In
$d=5$, the critical exponents $\gamma$ and $\Delta$ are significantly larger
than the mean-field values but no consistent evidence for the AT line is
found. Thus, our results are \textit{consistent}
with $6$ being the lower critical
dimension for the AT line. However it is also possible that an AT line 
\textit{does} occur for $d < 6$ but the series are too short to see it.

Finally we compare our results with other work.
The early renormalization group calculation of Bray and
Roberts~\cite{bray:80b} did not find a stable, perturbative fixed point
corresponding to an AT line below the upper critical dimension of 6. While
this result is consistent with there being no AT line below $d=6$ it is also
possible that a non-perturbative fixed point is present for this range of
dimension.  Other studies 
used Monte Carlo (MC) simulations on one-dimensional long-range 
interactions models~\cite{larson:13,baity-jesietal:14,banosetal:12}.
The results were analysed using finite-size scaling (FSS) theory. 
While all analyses found evidence for AT line for interactions corresponding to $d$ above $6$, different
conclusions were reached in lower dimensions~\cite{larson:13,baity-jesietal:14,banosetal:12}.

Our approach is complimentary to MC in that we study short-range models directly on
d-dimensional hypercubic lattices and that the series represent equilibrium property
of the \textit{infinite} system. Thus, the combined MC and series evidence provides a strong case
for an AT line in short-range models at least in high enough dimensions.
It would be challenging, but worthwhile, to try to
extend the series approach to include higher order terms.

\begin{acknowledgments}
One of us (APY) would like to thank the hospitality of the Indian Institute of
Science, Bangalore and the support of a DST-IISc Centenary Chair Professorship.
He is particularly grateful
for stimulating discussions with H.~Krishnamurthy which initiated this
project. The work of RRPS is supported in part by US NSF grant number DMR-1306048.
\end{acknowledgments}

\bibliography{refs}

\end{document}